\documentclass[proof]{WileyASNA-v1}

\articletype{Conference Paper}%

\received{xxxxx}
\revised{xxxxx}
\accepted{xxxxx}

\def \bea {\begin{eqnarray}}
\def \eea {\end{eqnarray}}
\def \nn   {\nonumber}
\begin{document}

\title{Consequences of Minimal Length Discretization on Line Element, Metric Tensor and Geodesic Equation}

\author[1,2]{Abdel Nasser Tawfik*}
\author[3]{Abdel Magied Diab}
\author[4]{Sameh Shenawy}
\author[5]{Eiman Abou El Dahab}

\address[1]{\orgdiv{Egyptian Center for Theoretical Physics (ECTP)}, \orgaddress{\state{12588 Giza}, \country{Egypt.}}}

\address[2]{\orgdiv{Goethe University}, \orgname{Institute for Theoretical Physics}, \orgaddress{\state{D-60438, Frankfurt am Main}, \country{Germany.}}}

\address[3]{\orgdiv{Modern University for Technology and Information (MTI)}, \orgname{Faculty of Engineering}, \orgaddress{\state{11571 Cairo}, \country{Egypt.}}}

\address[4]{\orgdiv{Modern Academy for Engineering}, \orgname{Basic Science Department}, \orgaddress{\state{11571 Cairo}, \country{Egypt}}}

\address[5]{\orgdiv{Modern University for Technology and Information (MTI)}, \orgname{Faculty of Computers and Information}, \orgaddress{\state{11671 Cairo}, \country{Egypt}}}

\corres{*\email{tawfik@itp.uni-frankfurt.de}}

\abstract{When minimal length uncertainty emerging from generalized uncertainty principle (GUP) is thoughtfully implemented, it is of great interest to consider its impacts on {\it ''gravitational} Einstein field equations (gEFE) and to try to find out whether consequential modifications in metric manifesting properties of quantum geometry due to quantum gravity. GUP takes into account the gravitational impacts on the noncommutation relations of length (distance) and momentum operators or time and energy operators, etc. On the other hand, gEFE relates {\it classical geometry or general relativity gravity} to the energy-momentum tensors, i.e. proposing quantum equations of state. Despite the technical difficulties, we confront GUP to the metric tensor so that the line element and the geodesic equation in flat and curved space are accordingly modified. The latter apparently encompasses acceleration, jerk, and snap (jounce) of a particle in the {\it ''quasi-quantized''} gravitational field. Finite higher-orders of acceleration apparently manifest phenomena such as accelerating expansion and transitions between different radii of curvature, etc.   
}



\fundingInfo{ECTP-2020-15 and WLCAPP-2020-15}

\maketitle

\footnotetext{\textbf{Abbreviations:} GR, general relativity; gEFE, {\it ''gravitational} Einstein field equations; QM, quantum mechanics; GUP, generalized uncertainty principle}

\section{Introduction}
\label{intro}

Formulation of a consistent theory for quantum gravity is {\it per se} an ultimate goal but unfortunately still one of the open questions in physics. There were various attempts to reconcile principles of the theory of general relativity in an entire quantum framework \cite{Rovelli:1989za,Donoghue:1994dn}. The quantum gravity is conjectured to add new elements into GR and QM. For the latter, we mention a modification of Heisenberg uncertainty principle due to gravitational effects, i.e. GUP \cite{Tawfik:2014zca,Tawfik:2015rva}. The present work focuses on the earlier, namely the possible modifications in line element, metric tensor, and geodesics. 

A modified length element and quantization of gravity were found from another perspective of GR \cite{Feoli:1999cn,Capozziello:2000gb,Bozza:2001qm}. An overview on these types of models
can be found in refs \cite{Hess:book,Hess:2020ssc} and in a review contribution of IWARA2020. An extended class of metric tensors that are functions of an internal vectory $^a(x)$ was also discussed in ref. \cite{Oliveira:1985cj}. A spin-1 massless field of gravitational origin in GR is then emerged. Regarding an equivalence principle in extended gravity and nongeodesic motion, a machian request was scratched \cite{Licata:2016qmz}, where a direct coupling between the Ricci curvature scalar and the matter Lagrangian can be fixed in cosmological observations.

It intends to tackle the long-standing fundamental problem that the Einstein field equations (EFE) relate {\it nonquantized} semi-Riemannian geometry characterized by Ricci and Einstein tensors, which are directly depending on the metric tensor, with the {\it full-quantized} energy-momentum tensor \cite{stephani_kramer_maccallum_hoenselaers_herlt_2003}. Our approach is based on implementing minimal length uncertainty obtained from generalized uncertainty principle (GUP), which in turn is inspired by string theory, doubly special relativity, and black hole physics \cite{Tawfik:2014zca,Tawfik:2015rva} and seems to be comparable to the Planck length, where fluctuations in {\it quasi-quantized} manifold likely emerge \cite{Tawfik:2016uhs}. We propose that this basic approach helps in characterizing the potential impacts of {\it full}-quantization on EFE. We also believe that this likely unveils the quantum nature of the cosmic geometry. We elaborate the corresponding modification in the line element, metric tensor, and geodesics and then show how this helps in manifesting properties of quantum geometry due to quantum gravity. The present study introduces a fundamental approach to the observations that the universe likely expands faster than the GR expectation. Unless forces deriving this kind of expansion, ingredients such as dark energy and cosmological constant might remain overdue \cite{Peebles:2002gy}.

\section{Generalized uncertainly principle and spacetime metric tensor}
\label{sec:gup}

Heisenberg uncertainty principle (HUP) dictates how to constrain the uncertainties in the quantum noncommutation relations of length (position) and momentum operators or of time and energy operators, for instance. On the other hand, when the gravitational influences are thoughtfully taken into account, a generalized uncertainty principle is then emerged so that an alternative quantum gravity approach for string theory, doubly special relativity, and black hole physics has been provided \cite{Tawfik:2014zca,Tawfik:2015rva}, e.g. a finite minimal length \cite{Kempf:1994su},
\bea
\Delta x\, \Delta p\geq \frac{\hbar}{2} \left[1+ \beta (\Delta p)^2 \right], \label{GUP}
\eea 
where $\Delta x$ and $\Delta p$ are length and momentum uncertainties, respectively. The GUP parameter $\beta = \beta_0 (\ell_p/\hbar)^2 = \beta_0/ (m_p c)^2$ with the Planck length $\ell_p=\sqrt{\hbar G/c^3}=1.977 \times 10^{-16}~$GeV$^{-1}$ and mass $m_p=\sqrt{\hbar c/G}= 1.22 \times 10^{19}~$GeV$/c^2$. Upper bounds on the dimensionless parameter $\beta_0$ should be put from astronomical observations, such as recent gravitational waves, $\beta_0 \lesssim 5.5 \times 10^{60}$. Equation (\ref{GUP}) seems to exhibit the existence of a minimum length uncertainty, 
\bea
\Delta x_{\mbox{min}} \approx \hbar \sqrt{\beta} &=& \ell_p \sqrt{\beta_0}.
\eea 

GUP also exhibits features of the UV/IR correspondence that $\Delta x$ increases rapidly (IR) as the $\Delta p$ grosses beyond the order of the Planck scale (UV) \cite{Maldacena:1997re,Gubser:1998bc,Witten:1998qj}. The UV/IR correspondence could be applied to various subjects of short vs. long distance physics, for instance, the {\it ''deformed''} commutation relations. We assume that the current problem of minimal length discretization would be solved by such a correspondence. 

Analogous to Eq. (\ref{GUP}), the canonical noncommutation relation of quantum operators of length and momentum reads 
\bea
\left[\hat{x}_i\,, \hat{p}_j\right] \geq \delta_{ij} i\, \hbar \left(1 + \beta p^2 \right), \label{GUP2}
\eea
where $p^2=g_{ij} p^{0i}\, p^{0j}$ and $g_{ij}$ is the Minkowski spacetime metric tensor, for instance $(-,+,+,+)$. The length and momentum operators, respectively, are defined as
\bea 
\hat{x}_i &=& \hat{x}_{0i} (1+\beta p^2),  \label{GUP3a}\\
\hat{p}_j &=& \hat{p}_{0j}, \label{GUP3b}
\eea
in which the operators $\hat{x}_{0i}$ and $\hat{p}_{0j}$ are to be derived from  the corresponding noncommunitation relation
\bea
[\hat{x}_{0i}, \hat{p}_{0j}]&=&\delta_{ij} i\, \hbar.
\eea 

\section{Spacetime geometry: line element and metric tensor}
\label{sec:st}

When including such quantum noncommutation operations, section \ref{sec:gup}, in the spacetime geometry, for instance EFE, the Minkowskian manifold of the line element 
\bea
ds^2 &=& g_{\mu\nu} dx^\mu dx^\nu, 
\eea
where $\mu$, $\nu$, and $\lambda=0, 1, 2, 3$, enlarges to an eight-dimensional spacetime tangent bundle similar to the one with the coordinates $x^A = (x^\mu(\zeta^a), \beta \dot{x}^\mu(\zeta^a))$, where $\dot{x}^\mu = dx^\mu/d\zeta^\mu$ \cite{Brandt:2000gka}. This leads to a modification in the line element, i.e. a new metric related the quantum geometry,
\bea
d \tilde{s}^{2} = g_{AB} \, dx^A \, dx^B, \label{eq:dS}
\eea
where $g_{AB} = g_{\mu_\nu}\otimes g_{\mu\nu}$. 

With some trivial approximations, the manifold can be reduced to the effective four-dimensional spacetime geometry, where $x^A= x^A(\zeta^\mu)$. Therefore, the modified four-dimensional metric tensor reads
 \bea
 \tilde{g}_{\mu\nu} &=& g_{AB} \frac{\partial x^A}{\partial \zeta^\mu} \frac{\partial x^B}{\partial \zeta^\nu} \simeq g_{ab} \Big[\frac{\partial x^a}{\partial \zeta^\mu}  \frac{\partial x^b}{\partial \zeta^\nu} + \beta \frac{\partial \dot{x}^a}{\partial \zeta^\mu}  \frac{\partial \dot{x}^b}{\partial \zeta^\nu} \Big]  \nn \\
 &\simeq&  \left(1+ \beta \ddot{x}^\lambda \ddot{x}_\lambda \right) g_{\mu\nu},
 \eea
where $\ddot{x}^\mu = \partial \dot{x}^\mu/\partial \zeta^\mu$ is the four-dimensional acceleration. The indices $A, B, a$, and $b$ run over $0, 1, \cdots, 7$. 
\begin{itemize}
\item For flat spacetime, where $g_{\mu \nu} = \eta_{\mu\nu}$, the modified metric tensor can be expressed es
\bea
\tilde{g}_{\mu\nu} &=&  \Big(1+ \beta \ddot{x}^\lambda \ddot{x}_\lambda \Big) \eta_{\mu \nu} =\eta_{\mu \nu} + h_{\mu\nu},
\eea
where $h_{\mu\nu} = \beta \ddot{x}^\lambda \ddot{x}_\lambda \eta_{\mu \nu} $ encompasses the quantum contributions to the spacetime geometry. 
\item In the limit that $ h\rightarrow 0$, the quantum corrections added entirely diminish and the {\it classical} EFE, i.e. GR EFE, can be restored. 
\end{itemize}
Thus, the principle of the general covariance is apparently satisfied also in the absence of the gravitational effects on the modified Minkowski metric. 

The modified four-dimensional line element is then expressed as
\bea
d \tilde{s}^{2} &=& g_{\mu\nu} \left( dx^\mu dx^\nu+ \beta^2  d\dot{x}^\mu \; d\dot{x}^\nu\right)\nn \\
&=& \left(1+ \beta^2 \ddot{x}^\lambda \ddot{x}_\lambda \right) ds^2.
\eea

\section{Spacetime geometry: geodesic equation}
\label{sec:ge}

Accordingly, the properties of the manifold in special and general relativity can also be generalized. We limit the discussion to modified geodesics, where the notion of a {\it ''straight line''} is generalized to curved spacetime. In this section, we propose a theory for the possible consequences of length discretization based on an approach to quantum gravity, GUP, on the world line of a free particle. With this regard, we recall that GR assumes gravity as a consequence of curved spacetime geometry and the {\it ''quantized''} energy-momentum tensor is the source of spacetime curvature. Our approach follows the same line with a major difference that the length is discretized and accordingly rhs of EFE.

By using the variational principle and by extremizing the path $s_{AB}$, the geodesic equation can be formulated. Due to the proposed length quantization, we get 
\begin{itemize}
\item for flat space 
\bea
\beta {\cal L} \frac{d^2 \dot{x}^{\mu}}{d \tau^2} - \frac{d x^{\mu}}{d \tau} + c &=& 0,
\eea
\item and for curved space
\bea
\frac{d^2 x^{2}}{d \tau^2} &-& \beta \frac{d}{d \tau} \left({\cal L} \frac{d^3 x^{2}}{d \tau^3}\right)  = \nn \\
&-& \Gamma^2_{\mu \nu} \frac{d x^{\mu}}{d \tau} \frac{d x^{nu}}{d \tau} +\beta g^{2\alpha} g_{\mu\nu, \alpha} \frac{d^2 x^{\mu}}{d^2 \tau} \frac{d^2 x^{nu}}{d \tau^2} \nn \\
&+& \beta {\cal L} g^{2\alpha} g_{\mu\alpha, \gamma} \left[
\frac{d x^{\gamma}}{d \tau} \frac{d^2 \dot{x}^{\mu}}{d \tau^2} + \frac{d}{d \tau} \left(\frac{d x^{\gamma}}{d \tau} \frac{d \dot{x}^{\mu}}{d \tau}\right)\right] \nn \\
&+& \beta {\cal L} g^{2\alpha} g_{\mu\alpha, \gamma, \delta}  \frac{d x^{\delta}}{d \tau} \frac{d x^{\gamma}}{d \tau} \frac{d \dot{x}^{\mu}}{d \tau}, \label{eq:geodsc1}
\eea
where 
\bea
\tau &=& \int {\cal L}(s, \dot{x}, \ddot{x})\, ds, \nn \\
\dot{x}^{\mu} &=& \frac{d x^{\mu}}{d s}, \nn \\
g_{\mu\nu, \alpha} &=& \frac{\partial g_{\mu \nu}}{\partial x^{\alpha}}, \nn \\
g_{\mu\alpha, \gamma, \delta} &=& \frac{\partial}{\partial x^{\delta}}  \left(\frac{\partial g_{\mu \alpha}}{\partial x^{\gamma}}\right), \nn \\
\Gamma^2_{\mu\nu} &=& \frac{1}{2} g^{2\alpha} \left[g_{\mu \alpha, \nu} - g_{\alpha \nu, \mu} + g_{\mu \nu, \alpha}\right], \nn \\
{\cal L} &=& \left[g_{\mu\nu} \left(
\frac{d x^{\mu}}{d s} \frac{d x^{\nu}}{d s} + \beta \frac{d \dot{x}^{\mu}}{d s} \frac{d \dot{x}^{\nu}}{d s} \right)\right]^{1/2}. \nn
\eea
\end{itemize}
The $\beta$-terms appeared in Eq. (\ref{eq:geodsc1}) distinguish this expression from the GR geodesics. In the section that follows, we discuss on these terms, where the consequences of the minimal length discretization are integrated in and summarize our final conclusions.

\section{Conclusions}

It is worthwhile to highlight that Christoffel connection and the equivalence principle are not affected by the minimal length uncertainty. Thus, we conclude that GUP reconciles with the equivalence principle and simultaneously the equivalence principle is not violated. In such a way, this gives possibilities to recover the violation of the equivalence principle in presence of the quantum gravity through the quantum geometry characterized by minimal length discretization. The reason that this essential result apparently contradicts refs. \cite{Ghosh:2013qra, Scardigli:2007bw}, can be understood that we have implemented GUP to the basic metric tensor. Accordingly, we have obtained a modified metric tensor and line element. Both are very fundamentals of the proposed geometry. Having both quantities modified, we could derive the corresponding geodesics. The appearance of vibration and/or sudden transition, as shall be discussed shortly, is apparently stemming from the {\it quasi-quantized} geometry or the proposed approach to quantum gravity, where the curvature emerged by the quantized energy-momentum tensor seems gains corrections, as well. 

As outlined in section \ref{sec:st}, the corrections to the line element are combined in the term $\beta^2 \ddot{x}^\lambda \ddot{x}_\lambda$, which manifest the essential contributions added to by acceleration of the expansion. In other words, it seems that the length discretization as guaranteed by GUP emphasizes that even the line element wouldn't only expand but also accelerates. It would be noticed that even if the GUP parameter $\beta$ squared would assure that this factor remains small, the product $\ddot{x}^\lambda \ddot{x}_\lambda$ raises the values of this correction term. In the same matter, the metric tensor obtains corrections, $\beta \ddot{x}^\lambda \ddot{x}_\lambda$, as well. Here, the factor to the product $\ddot{x}^\lambda \ddot{x}_\lambda$ is simply the GUP factor $\beta$.

Even if classical GR necessarily invokes metric tensor in four dimensions, we assume that the corresponding reference frames experience different expansions in time and space. The metric expansion proposed in the present work are related to changes in metric tensor with time. Although the possible differences in temporal and spacial expansions, we straightforwardly impose the origin of an accelerated expansion. The spacetime seems to shrink or grow as the corresponding geodesics converges or diverges, where the length discretization plays the major role.

We also notice that the geodesics, Eq. (\ref{eq:geodsc1}), is not only providing the acceleration of a particle in a gravitational field, but higher-order derivatives, as well \cite{Eager_2016}, namely the snap or jounce, $x^{(4)}$, which in turn is derived from the jerk, $x^{(3)}$. The jerk gives the change in the force acting on that particle, while the snap is resulted from change in the jerk, itself. That both quantities are finite means that vibration or sudden transitions would occur between different radii of the curvature. Acceleration, as in GR geodesics, without jerk is just a static load, i.e neither vibration nor transition are allowed. 

Due GUP and the corresponding minimal length descretization, the corrections added to the line element, metric tensor, and geodesics emphasize that evolution of the universe as theorized by classical GR is also accelerated \cite{Riess:1998cb,Velten:2017ire}. While the line element and metric tensor get additional terms of acceleration products, $\ddot{x}^\lambda \ddot{x}_\lambda$, the geodesics  on the other hand is corrected with higher order acceleration derivatives, such as snap $x^{(4)}$ and jerk $x^{(3)}$.

\section*{Acknowledgments}

The authors are very grateful to organizers of the $9^{\mbox{th}}$ International Workshop on Astronomy and Relativistic Astrophysics (IWARA2020 Video Conference) for their kind invitation.  
  
\bibliography{ATawfikRefsIWARA2020}%

\begin{thebibliography}{}

\bibitem [\protect \citeauthoryear {%
Bozza%
, Feoli%
, Lambiase%
, Papini%
\BCBL {}\ \BBA {} Scarpetta%
}{%
Bozza%
\ \protect \BOthers {.}}{%
{\protect \APACyear {2001}}%
}]{%
Bozza:2001qm}
\APACinsertmetastar {%
Bozza:2001qm}%
\begin{APACrefauthors}%
Bozza, V.%
, Feoli, A.%
, Lambiase, G.%
, Papini, G.%
\BCBL {}\ \BBA {} Scarpetta, G.%
\end{APACrefauthors}%
\unskip\
\newblock
\APACrefYearMonthDay{2001}{}{},
\newblock
\unskip
\newblock
\APACjournalVolNumPages{Phys. Lett. A}{283}{}{53--61}.
\newblock
\begin{APACrefDOI} \doi{10.1016/S0375-9601(01)00230-4} \end{APACrefDOI}
\PrintBackRefs{\CurrentBib}

\bibitem [\protect \citeauthoryear {%
Brandt%
}{%
Brandt%
}{%
{\protect \APACyear {2000}}%
}]{%
Brandt:2000gka}
\APACinsertmetastar {%
Brandt:2000gka}%
\begin{APACrefauthors}%
Brandt, H\BPBI E.%
\end{APACrefauthors}%
\unskip\
\newblock
\APACrefYearMonthDay{2000}{}{},
\newblock
\unskip
\newblock
\APACjournalVolNumPages{Found. Phys. Lett.}{13}{6}{581--588}.
\newblock
\begin{APACrefDOI} \doi{10.1023/A:1007862431369} \end{APACrefDOI}
\PrintBackRefs{\CurrentBib}

\bibitem [\protect \citeauthoryear {%
Capozziello%
, Feoli%
, Lambiase%
, Papini%
\BCBL {}\ \BBA {} Scarpetta%
}{%
Capozziello%
\ \protect \BOthers {.}}{%
{\protect \APACyear {2000}}%
}]{%
Capozziello:2000gb}
\APACinsertmetastar {%
Capozziello:2000gb}%
\begin{APACrefauthors}%
Capozziello, S.%
, Feoli, A.%
, Lambiase, G.%
, Papini, G.%
\BCBL {}\ \BBA {} Scarpetta, G.%
\end{APACrefauthors}%
\unskip\
\newblock
\APACrefYearMonthDay{2000}{}{},
\newblock
\unskip
\newblock
\APACjournalVolNumPages{Phys. Lett. A}{268}{}{247--254}.
\newblock
\begin{APACrefDOI} \doi{10.1016/S0375-9601(00)00215-2} \end{APACrefDOI}
\PrintBackRefs{\CurrentBib}

\bibitem [\protect \citeauthoryear {%
Donoghue%
}{%
Donoghue%
}{%
{\protect \APACyear {1994}}%
}]{%
Donoghue:1994dn}
\APACinsertmetastar {%
Donoghue:1994dn}%
\begin{APACrefauthors}%
Donoghue, J\BPBI F.%
\end{APACrefauthors}%
\unskip\
\newblock
\APACrefYearMonthDay{1994}{}{},
\newblock
\unskip
\newblock
\APACjournalVolNumPages{Phys. Rev. D}{50}{}{3874--3888}.
\newblock
\begin{APACrefDOI} \doi{10.1103/PhysRevD.50.3874} \end{APACrefDOI}
\PrintBackRefs{\CurrentBib}

\bibitem [\protect \citeauthoryear {%
Eager%
, Pendrill%
\BCBL {}\ \BBA {} Reistad%
}{%
Eager%
\ \protect \BOthers {.}}{%
{\protect \APACyear {2016}}%
}]{%
Eager_2016}
\APACinsertmetastar {%
Eager_2016}%
\begin{APACrefauthors}%
Eager, D.%
, Pendrill, A\BHBI M.%
\BCBL {}\ \BBA {} Reistad, N.%
\end{APACrefauthors}%
\unskip\
\newblock
\APACrefYearMonthDay{2016}{oct}{},
\newblock
\unskip
\newblock
\APACjournalVolNumPages{European Journal of Physics}{37}{6}{065008}.
\newblock
\begin{APACrefDOI} \doi{10.1088/0143-0807/37/6/065008} \end{APACrefDOI}
\PrintBackRefs{\CurrentBib}

\bibitem [\protect \citeauthoryear {%
Feoli%
, Lambiase%
, Papini%
\BCBL {}\ \BBA {} Scarpetta%
}{%
Feoli%
\ \protect \BOthers {.}}{%
{\protect \APACyear {1999}}%
}]{%
Feoli:1999cn}
\APACinsertmetastar {%
Feoli:1999cn}%
\begin{APACrefauthors}%
Feoli, A.%
, Lambiase, G.%
, Papini, G.%
\BCBL {}\ \BBA {} Scarpetta, G.%
\end{APACrefauthors}%
\unskip\
\newblock
\APACrefYearMonthDay{1999}{}{},
\newblock
\unskip
\newblock
\APACjournalVolNumPages{Phys. Lett. A}{263}{}{147--153}.
\newblock
\begin{APACrefDOI} \doi{10.1016/S0375-9601(99)00706-9} \end{APACrefDOI}
\PrintBackRefs{\CurrentBib}

\bibitem [\protect \citeauthoryear {%
Ghosh%
}{%
Ghosh%
}{%
{\protect \APACyear {2014}}%
}]{%
Ghosh:2013qra}
\APACinsertmetastar {%
Ghosh:2013qra}%
\begin{APACrefauthors}%
Ghosh, S.%
\end{APACrefauthors}%
\unskip\
\newblock
\APACrefYearMonthDay{2014}{}{},
\newblock
\unskip
\newblock
\APACjournalVolNumPages{Class. Quant. Grav.}{31}{}{025025}.
\newblock
\begin{APACrefDOI} \doi{10.1088/0264-9381/31/2/025025} \end{APACrefDOI}
\PrintBackRefs{\CurrentBib}

\bibitem [\protect \citeauthoryear {%
Gubser%
, Klebanov%
\BCBL {}\ \BBA {} Polyakov%
}{%
Gubser%
\ \protect \BOthers {.}}{%
{\protect \APACyear {1998}}%
}]{%
Gubser:1998bc}
\APACinsertmetastar {%
Gubser:1998bc}%
\begin{APACrefauthors}%
Gubser, S.%
, Klebanov, I\BPBI R.%
\BCBL {}\ \BBA {} Polyakov, A\BPBI M.%
\end{APACrefauthors}%
\unskip\
\newblock
\APACrefYearMonthDay{1998}{}{},
\newblock
\unskip
\newblock
\APACjournalVolNumPages{Phys. Lett. B}{428}{}{105--114}.
\newblock
\begin{APACrefDOI} \doi{10.1016/S0370-2693(98)00377-3} \end{APACrefDOI}
\PrintBackRefs{\CurrentBib}

\bibitem [\protect \citeauthoryear {%
P.~Hess%
}{%
P.~Hess%
}{%
{\protect \APACyear {2020}}%
}]{%
Hess:2020ssc}
\APACinsertmetastar {%
Hess:2020ssc}%
\begin{APACrefauthors}%
Hess, P.%
\end{APACrefauthors}%
\unskip\
\newblock
\APACrefYearMonthDay{2020}{}{},
\newblock
\unskip
\newblock
\APACjournalVolNumPages{Prog. Part. Nucl. Phys.}{114}{}{103809}.
\newblock
\begin{APACrefDOI} \doi{10.1016/j.ppnp.2020.103809} \end{APACrefDOI}
\PrintBackRefs{\CurrentBib}

\bibitem [\protect \citeauthoryear {%
P\BPBI O.~Hess%
, Schafer%
\BCBL {}\ \BBA {} Greiner%
}{%
P\BPBI O.~Hess%
\ \protect \BOthers {.}}{%
{\protect \APACyear {2015}}%
}]{%
Hess:book}
\APACinsertmetastar {%
Hess:book}%
\begin{APACrefauthors}%
Hess, P\BPBI O.%
, Schafer, M.%
\BCBL {}\ \BBA {} Greiner, W.%
\end{APACrefauthors}%
\unskip\
\newblock
\APACrefYear{2015},
\newblock
\APACrefbtitle {Pseudo-complex General Relativity} {Pseudo-complex General
  Relativity}.
\newblock
\APACaddressPublisher{Heidelberg}{Springer}.
\PrintBackRefs{\CurrentBib}

\bibitem [\protect \citeauthoryear {%
Kempf%
, Mangano%
\BCBL {}\ \BBA {} Mann%
}{%
Kempf%
\ \protect \BOthers {.}}{%
{\protect \APACyear {1995}}%
}]{%
Kempf:1994su}
\APACinsertmetastar {%
Kempf:1994su}%
\begin{APACrefauthors}%
Kempf, A.%
, Mangano, G.%
\BCBL {}\ \BBA {} Mann, R\BPBI B.%
\end{APACrefauthors}%
\unskip\
\newblock
\APACrefYearMonthDay{1995}{}{},
\newblock
\unskip
\newblock
\APACjournalVolNumPages{Phys. Rev. D}{52}{}{1108--1118}.
\newblock
\begin{APACrefDOI} \doi{10.1103/PhysRevD.52.1108} \end{APACrefDOI}
\PrintBackRefs{\CurrentBib}

\bibitem [\protect \citeauthoryear {%
Licata%
, Corda%
\BCBL {}\ \BBA {} Benedetto%
}{%
Licata%
\ \protect \BOthers {.}}{%
{\protect \APACyear {2016}}%
}]{%
Licata:2016qmz}
\APACinsertmetastar {%
Licata:2016qmz}%
\begin{APACrefauthors}%
Licata, I.%
, Corda, C.%
\BCBL {}\ \BBA {} Benedetto, E.%
\end{APACrefauthors}%
\unskip\
\newblock
\APACrefYearMonthDay{2016}{}{},
\newblock
\unskip
\newblock
\APACjournalVolNumPages{Grav. Cosmol.}{22}{1}{48--53}.
\newblock
\begin{APACrefDOI} \doi{10.1134/S0202289316010102} \end{APACrefDOI}
\PrintBackRefs{\CurrentBib}

\bibitem [\protect \citeauthoryear {%
Maldacena%
}{%
Maldacena%
}{%
{\protect \APACyear {1999}}%
}]{%
Maldacena:1997re}
\APACinsertmetastar {%
Maldacena:1997re}%
\begin{APACrefauthors}%
Maldacena, J\BPBI M.%
\end{APACrefauthors}%
\unskip\
\newblock
\APACrefYearMonthDay{1999}{}{},
\newblock
\unskip
\newblock
\APACjournalVolNumPages{Int. J. Theor. Phys.}{38}{}{1113--1133}.
\newblock
\begin{APACrefDOI} \doi{10.1023/A:1026654312961} \end{APACrefDOI}
\PrintBackRefs{\CurrentBib}

\bibitem [\protect \citeauthoryear {%
Oliveira%
}{%
Oliveira%
}{%
{\protect \APACyear {1985}}%
}]{%
Oliveira:1985cj}
\APACinsertmetastar {%
Oliveira:1985cj}%
\begin{APACrefauthors}%
Oliveira, C\BPBI G.%
\end{APACrefauthors}%
\unskip\
\newblock
\APACrefYearMonthDay{1985}{}{},
\newblock
\unskip
\newblock
\APACjournalVolNumPages{Int. J. Theor. Phys.}{24}{}{1081}.
\newblock
\begin{APACrefDOI} \doi{10.1007/BF00671307} \end{APACrefDOI}
\PrintBackRefs{\CurrentBib}

\bibitem [\protect \citeauthoryear {%
Peebles%
\ \BBA {} Ratra%
}{%
Peebles%
\ \BBA {} Ratra%
}{%
{\protect \APACyear {2003}}%
}]{%
Peebles:2002gy}
\APACinsertmetastar {%
Peebles:2002gy}%
\begin{APACrefauthors}%
Peebles, P.%
\BCBT {}\ \BBA {} Ratra, B.%
\end{APACrefauthors}%
\unskip\
\newblock
\APACrefYearMonthDay{2003}{}{},
\newblock
\unskip
\newblock
\APACjournalVolNumPages{Rev. Mod. Phys.}{75}{}{559--606}.
\newblock
\begin{APACrefDOI} \doi{10.1103/RevModPhys.75.559} \end{APACrefDOI}
\PrintBackRefs{\CurrentBib}

\bibitem [\protect \citeauthoryear {%
Riess%
\ \protect \BOthers {.}}{%
Riess%
\ \protect \BOthers {.}}{%
{\protect \APACyear {1998}}%
}]{%
Riess:1998cb}
\APACinsertmetastar {%
Riess:1998cb}%
\begin{APACrefauthors}%
Riess, A\BPBI G.%
\BCBT {}\ \BOthersPeriod {.}
\end{APACrefauthors}%
\unskip\
\newblock
\APACrefYearMonthDay{1998}{}{},
\newblock
\unskip
\newblock
\APACjournalVolNumPages{Astron. J.}{116}{}{1009--1038}.
\newblock
\begin{APACrefDOI} \doi{10.1086/300499} \end{APACrefDOI}
\PrintBackRefs{\CurrentBib}

\bibitem [\protect \citeauthoryear {%
Rovelli%
\ \BBA {} Smolin%
}{%
Rovelli%
\ \BBA {} Smolin%
}{%
{\protect \APACyear {1990}}%
}]{%
Rovelli:1989za}
\APACinsertmetastar {%
Rovelli:1989za}%
\begin{APACrefauthors}%
Rovelli, C.%
\BCBT {}\ \BBA {} Smolin, L.%
\end{APACrefauthors}%
\unskip\
\newblock
\APACrefYearMonthDay{1990}{}{},
\newblock
\unskip
\newblock
\APACjournalVolNumPages{Nucl. Phys. B}{331}{}{80--152}.
\newblock
\begin{APACrefDOI} \doi{10.1016/0550-3213(90)90019-A} \end{APACrefDOI}
\PrintBackRefs{\CurrentBib}

\bibitem [\protect \citeauthoryear {%
Scardigli%
\ \BBA {} Casadio%
}{%
Scardigli%
\ \BBA {} Casadio%
}{%
{\protect \APACyear {2009}}%
}]{%
Scardigli:2007bw}
\APACinsertmetastar {%
Scardigli:2007bw}%
\begin{APACrefauthors}%
Scardigli, F.%
\BCBT {}\ \BBA {} Casadio, R.%
\end{APACrefauthors}%
\unskip\
\newblock
\APACrefYearMonthDay{2009}{}{},
\newblock
\unskip
\newblock
\APACjournalVolNumPages{Int. J. Mod. Phys. D}{18}{}{319--327}.
\newblock
\begin{APACrefDOI} \doi{10.1142/S0218271809014455} \end{APACrefDOI}
\PrintBackRefs{\CurrentBib}

\bibitem [\protect \citeauthoryear {%
Stephani%
, Kramer%
, MacCallum%
, Hoenselaers%
\BCBL {}\ \BBA {} Herlt%
}{%
Stephani%
\ \protect \BOthers {.}}{%
{\protect \APACyear {2003}}%
}]{%
stephani_kramer_maccallum_hoenselaers_herlt_2003}
\APACinsertmetastar {%
stephani_kramer_maccallum_hoenselaers_herlt_2003}%
\begin{APACrefauthors}%
Stephani, H.%
, Kramer, D.%
, MacCallum, M.%
, Hoenselaers, C.%
\BCBL {}\ \BBA {} Herlt, E.%
\end{APACrefauthors}%
\unskip\
\newblock
\APACrefYearMonthDay{2003}{}{},
\newblock
{\BBOQ}\APACrefatitle {Classification of the Ricci tensor and the
  energy-momentum tensor} {Classification of the Ricci tensor and the
  energy-momentum tensor}.{\BBCQ}
\newblock
\BIn{} \APACrefbtitle {Exact Solutions of Einstein's Field Equations} {Exact
  Solutions of Einstein's Field Equations}\ \PrintOrdinal{2}\ \BEd,
  \BPG~57–67.
\newblock
\APACaddressPublisher{}{Cambridge University Press}.
\newblock
\begin{APACrefDOI} \doi{10.1017/CBO9780511535185.007} \end{APACrefDOI}
\PrintBackRefs{\CurrentBib}

\bibitem [\protect \citeauthoryear {%
A.~Tawfik%
\ \BBA {} Diab%
}{%
A.~Tawfik%
\ \BBA {} Diab%
}{%
{\protect \APACyear {2016}}%
}]{%
Tawfik:2016uhs}
\APACinsertmetastar {%
Tawfik:2016uhs}%
\begin{APACrefauthors}%
Tawfik, A.%
\BCBT {}\ \BBA {} Diab, A.%
\end{APACrefauthors}%
\unskip\
\newblock
\APACrefYearMonthDay{2016}{}{},
\newblock
\unskip
\newblock
\APACjournalVolNumPages{Indian J. Phys.}{90}{10}{1095--1103}.
\newblock
\begin{APACrefDOI} \doi{10.1007/s12648-016-0855-4} \end{APACrefDOI}
\PrintBackRefs{\CurrentBib}

\bibitem [\protect \citeauthoryear {%
A\BPBI N.~Tawfik%
\ \BBA {} Diab%
}{%
A\BPBI N.~Tawfik%
\ \BBA {} Diab%
}{%
{\protect \APACyear {2014}}%
}]{%
Tawfik:2014zca}
\APACinsertmetastar {%
Tawfik:2014zca}%
\begin{APACrefauthors}%
Tawfik, A\BPBI N.%
\BCBT {}\ \BBA {} Diab, A\BPBI M.%
\end{APACrefauthors}%
\unskip\
\newblock
\APACrefYearMonthDay{2014}{}{},
\newblock
\unskip
\newblock
\APACjournalVolNumPages{Int. J. Mod. Phys. D}{23}{12}{1430025}.
\newblock
\begin{APACrefDOI} \doi{10.1142/S0218271814300250} \end{APACrefDOI}
\PrintBackRefs{\CurrentBib}

\bibitem [\protect \citeauthoryear {%
A\BPBI N.~Tawfik%
\ \BBA {} Diab%
}{%
A\BPBI N.~Tawfik%
\ \BBA {} Diab%
}{%
{\protect \APACyear {2015}}%
}]{%
Tawfik:2015rva}
\APACinsertmetastar {%
Tawfik:2015rva}%
\begin{APACrefauthors}%
Tawfik, A\BPBI N.%
\BCBT {}\ \BBA {} Diab, A\BPBI M.%
\end{APACrefauthors}%
\unskip\
\newblock
\APACrefYearMonthDay{2015}{}{},
\newblock
\unskip
\newblock
\APACjournalVolNumPages{Rept. Prog. Phys.}{78}{}{126001}.
\newblock
\begin{APACrefDOI} \doi{10.1088/0034-4885/78/12/126001} \end{APACrefDOI}
\PrintBackRefs{\CurrentBib}

\bibitem [\protect \citeauthoryear {%
Velten%
, Gomes%
\BCBL {}\ \BBA {} Busti%
}{%
Velten%
\ \protect \BOthers {.}}{%
{\protect \APACyear {2018}}%
}]{%
Velten:2017ire}
\APACinsertmetastar {%
Velten:2017ire}%
\begin{APACrefauthors}%
Velten, H.%
, Gomes, S.%
\BCBL {}\ \BBA {} Busti, V\BPBI C.%
\end{APACrefauthors}%
\unskip\
\newblock
\APACrefYearMonthDay{2018}{}{},
\newblock
\unskip
\newblock
\APACjournalVolNumPages{Phys. Rev. D}{97}{8}{083516}.
\newblock
\begin{APACrefDOI} \doi{10.1103/PhysRevD.97.083516} \end{APACrefDOI}
\PrintBackRefs{\CurrentBib}

\bibitem [\protect \citeauthoryear {%
Witten%
}{%
Witten%
}{%
{\protect \APACyear {1998}}%
}]{%
Witten:1998qj}
\APACinsertmetastar {%
Witten:1998qj}%
\begin{APACrefauthors}%
Witten, E.%
\end{APACrefauthors}%
\unskip\
\newblock
\APACrefYearMonthDay{1998}{}{},
\newblock
\unskip
\newblock
\APACjournalVolNumPages{Adv. Theor. Math. Phys.}{2}{}{253--291}.
\newblock
\begin{APACrefDOI} \doi{10.4310/ATMP.1998.v2.n2.a2} \end{APACrefDOI}
\PrintBackRefs{\CurrentBib}

\end{thebibliography}

\end{document}